\documentclass[twocolumn,twoside,slac_two]{revtex4}
\usepackage{graphicx}
\usepackage{fancyhdr}
\usepackage{epsfig}
\usepackage{natbib}
\setcitestyle{numbers}
\usepackage{amsmath}
\usepackage{amssymb}
\def\rQCED{{\rm QCED}}
\begin{document}
\title{HERWIRI1.0(31): MC Realization of IR-Improvement for DGLAP-CS Parton Showers}


\author{B. F. L. Ward}
\affiliation{Department of Physics, Baylor University, Waco, TX 76798, USA}
\author{S. Joseph}
\affiliation{Department of Physics, Baylor University, Waco, TX 76798, USA}
\author{S. Majhi}
\affiliation{Theory Division, Saha Institute of Nuclear Physics, Kolkata 700 064, India}
\author{S. A. Yost}
\affiliation{Department of Physics, The Citadel, Charleston, SC 29409, USA}

\begin{abstract}
We have implemented the new IR-improved 
Dokshitzer-Gribov-Lipatov-Altarelli-Parisi-Callan-Symanzik
(DGLAP-CS) kernels recently developed by one of us in
the HERWIG6.5 environment to generate a new MC, HERWIRI1.0(31),
for hadron-hadron scattering at high energies.
We present MC data that illustrate 
the comparison between the parton shower generated by the 
standard DGLAP-CS kernels and that generated by the new IR-improved 
DGLAP-CS kernels. We also show comparisons with FNAL data
and we discuss possible 
implications for LHC phenomenology.
%This document serves as a template for the proceedings of the DPF-2009 conference.  
%Authors should prepare their papers using a copy and follow the guidelines described here.
%Please do not modify the page layout or styles. 
\end{abstract}

\maketitle

\thispagestyle{fancy}


\section{\label{intro} Introduction}

In the LHC era of precision QCD, which entails  
predictions for QCD processes at the total precision~\citep{jadach1} tag~\footnote{By total precision of a theoretical prediction we mean the technical and physical precisions combined in quadrature or otherwise as appropriate.} of $1\%$ or better, we need resummed ${\cal O}(\alpha_s^2L^n),\;{\cal O}(\alpha_s\alpha L^{n'}),\;{\cal O}(\alpha^2 L^{n''}), n = 0,1,2,\; n' = 0,1,2, n'' = 1,2$ corrections, in the presence of parton showers, on an event-by-event basis, without double counting and with exact phase space. The roles of QED and EW effects~\citep{qedeffects,radcor-ew} are integral parts of the discussion with which we 
deal by the simultaneous
resummation of QED and QCD large infrared(IR) 
effects, $QED\otimes QCD$ resummation
~\citep{qced} in the presence of parton showers, to be realized on an 
event-by-event basis by MC methods; for, as shown in
Refs.~\citep{radcor-ew}, no precision prediction for a hard LHC process
at the 1\% level can be complete without taking the large EW corrections into account. \par
In what follows, we first review
our approach to resummation and its relationship to those in Refs.~\citep{cattrent,scet}. Section 3 contains a summary of the attendant new IR-improved DGLAP-CS~\citep{dglap,cs} theory~\citep{irdglap1,irdglap2}.
Section 4 presents the implementation of the new IR-improved kernels in the framework of HERWIG6.5~\citep{herwig} to arrive at the new, IR-improved parton shower MC
HERWIRI1.0. We illustrate the effects of the IR-improvement first with the 
generic 2$\rightarrow$2 processes at LHC energies and then  with the specific
single Z production process at LHC energies. We compare with recent 
data from FNAL to make
direct contact with observation. 
Section~\ref{concl} contains our summary remarks.
\par%\vskip0.2cm
For reference purposes, we call attention to the 
analyses in 
Refs.~\citep{scott1,scott2}, wherein the authors have argued that
the current state-of-the-art theoretical precision tag on single Z
production at the LHC is $(4.1\pm0.3)\%=(1.51\pm 0.75)\%(QCD)\oplus 3.79(PDF)\oplus 0.38\pm 0.26(EW)\% $ and that the analogous estimate for single W production is $\sim 5.7$\%. One cannot emphasize too much that these 
estimates show how
much work is still needed to achieve the desired 1.0\% total precision tag
on the two respective processes, for example.\par
%,
%where the results of Refs.~\cite{cteq,mrst,mcnlo,fewz,resbos,horace,photos} have been
%used.
%\footnote{Recently, the 
%analogous estimate for single W production has been given ~\cite{scott2} as $\sim 5.7$\%.}\par%\vskip0.2cm
%%%Start Here
\section{QED$\otimes$QCD Resummation}
We make use of the discussion in Refs.~\citep{qced,irdglap1,irdglap2}, wherein
we have derived the following expression for the 
hard cross sections in the SM $SU_{2L}\times U_1\times SU_3^c$ EW-QCD theory{\small
\begin{eqnarray}
&d\hat\sigma_{\rm exp} = e^{\rm SUM_{IR}(QCED)}
   \sum_{{n,m}=0}^\infty\frac{1}{n!m!}\int\prod_{j_1=1}^n\frac{d^3k_{j_1}}{k_{j_1}} \cr
&\prod_{j_2=1}^m\frac{d^3{k'}_{j_2}}{{k'}_{j_2}}
\int\frac{d^4y}{(2\pi)^4}e^{iy\cdot(p_1+q_1-p_2-q_2-\sum k_{j_1}-\sum {k'}_{j_2})+
D_\rQCED} \cr
&\tilde{\bar\beta}_{n,m}(k_1,\ldots,k_n;k'_1,\ldots,k'_m)\frac{d^3p_2}{p_2^{\,0}}\frac{d^3q_2}{q_2^{\,0}},
%\end{split}
\label{subp15b}
\end{eqnarray}}\noindent
where the new YFS-style~\citep{yfs} residuals
$\tilde{\bar\beta}_{n,m}(k_1,\ldots,k_n;k'_1,\ldots,k'_m)$ have $n$ hard gluons and $m$ hard photons and we show the final state with two hard final
partons with momenta $p_2,\; q_2$ specified for a generic $2f$ final state for
definiteness. The infrared functions ${\rm SUM_{IR}(QCED)},\; D_\rQCED\; $
are defined in Refs.~\citep{qced,irdglap1,irdglap2}. This is the exact, 
simultaneous resummation of QED and QCD large IR effects.\par%\vskip0.2cm
One can see that our approach to 
QCD resummation is fully consistent with that of
Refs.~\citep{cattrent,scet} as follows. First, Ref.~\citep{geor1} has shown that the latter two approaches are equivalent. 
By using the color-spin density matrix realization of our residuals, we 
show in Refs.~\citep{irdglap1,irdglap2}
that our approach is consistent with that of Refs.~\citep{cattrent}
by exhibiting the transformation prescription from the resummation formula
for the theory in Refs.~\citep{cattrent} for the generic $2\rightarrow n$ parton process as given in Ref.~\citep{madg} to our theory as given for QCD by restricting (\ref{subp15b}) to its QCD component. In this way,
we capture the respective full quantum mechanical color-spin correlations in the results in Ref.~\citep{madg}.\par%\vskip0.2cm
\section{IR-Improved DGLAP-CS Theory}
We show in Refs.~\citep{irdglap1,irdglap2} that the result (\ref{subp15b})
allows us to improve in the IR regime \footnote{This 
should be distinguished from the also important
resummation in parton density evolution for the ``$z\rightarrow 0$'' regime,
where Regge asymptotics obtain -- see for example Ref.~\citep{ermlv,guido}. This
improvement must also be taken into account for precision LHC predictions.} 
the kernels in DGLAP-CS~\citep{dglap,cs}
theory as follows, using a standard notation:{\small
\begin{align}
P^{exp}_{qq}(z)&= C_F F_{YFS}(\gamma_q)e^{\frac{1}{2}\delta_q}\big[\frac{1+z^2}{1-z}(1-z)^{\gamma_q} \nonumber\\
&\qquad -f_q(\gamma_q)\delta(1-z)\big],\nonumber\\
P^{exp}_{Gq}(z)&= C_F F_{YFS}(\gamma_q)e^{\frac{1}{2}\delta_q}\frac{1+(1-z)^2}{z} z^{\gamma_q},\nonumber\\
P^{exp}_{GG}(z)&= 2C_G F_{YFS}(\gamma_G)e^{\frac{1}{2}\delta_G}\{ \frac{1-z}{z}z^{\gamma_G}+\frac{z}{1-z}(1-z)^{\gamma_G}\nonumber\\
&\qquad +\frac{1}{2}(z^{1+\gamma_G}(1-z)+z(1-z)^{1+\gamma_G})\nonumber\\
&\qquad  - f_G(\gamma_G) \delta(1-z)\},\nonumber\\
P^{exp}_{qG}(z)&= F_{YFS}(\gamma_G)e^{\frac{1}{2}\delta_G}\frac{1}{2}\{ z^2(1-z)^{\gamma_G}+(1-z)^2z^{\gamma_G}\},
%P_{qG}(z)&=\frac{1}{2}(z^2+(1-z)^2).
\label{dglap19}
\end{align}}
where the superscript ``exp'' indicates that the kernel has been resummed as
predicted by (\ref{subp15b}) when it is restricted to QCD alone -- see Refs.~\citep{irdglap1,irdglap2} for the corresponding details.
These results have been implemented by 
MC methods as we exhibit in what follows.\par%\vskip0.2cm
%\section{Illustrative  Results/Implications}
Let us first note that
a number of illustrative results and implications of the new 
kernels have been presented in Refs.~\citep{irdglap1,irdglap2,irdglap3-plb}.
Here, we call attention to the new scheme~\citep{irdglap2} which we now have 
%Firstly, we note that the connection to the higher order kernels in Refs.~\cite{high-ord-krnls} is done~\cite{irdglap1}. This opens the way for the 
%systematic improvement of the results presented herein accordingly.
%Secondly, in the NS case, we find~\cite{irdglap1} that the n=2 moment
%is modified by $\sim 5\%$ when evolved with (\ref{dglap19}) 
%from $2$GeV to $100$GeV with $n_f=5$
%and $\Lambda_{QCD}\cong .2GeV$, for illustration. This is thus relevant
%to the expected precision of the HERA final data analysis~\cite{hera-dat}.
%Thirdly, 
%we have been able to use
%(\ref{subp15b}) to resolve violation~\cite{sac-no-go,cat1} 
%of Bloch-Nordsieck cancellation in 
%ISR(initial state radiation) 
%at ${\cal O}(\alpha_s^2)$ for massive quarks~\cite{qmass-bw}.
%This opens the way to use the realistic quark masses as we introduce the
%higher order EW corrections in the presence of higher order QCD corrections -- note that the radiation probability in QED at the hard scale $Q$ involves the logarithm $\ln(Q^2/m_q^2)$ and it will not do to set $m_q=0$ to analyze these effects in a fully exclusive, differential even-by-event calculation of the type that we are constructing. 
%Fourthly, the threshold resummation implied by (\ref{subp15b}) for single Z
%production at LHC shows a $0.3\%$ QED effect and agrees with known exact
%results in QCD -- see Refs.~\cite{qced,baurall,exactqcd}. Fifthly, 
%we have a new scheme~\cite{irdglap2} 
for precision LHC theory: in an obvious notation,
\begin{equation}
\begin{split}
\sigma &=\sum_{i,j}\int dx_1dx_2F_i(x_1)F_j(x_2)\hat\sigma(x_1x_2s)\nonumber\\
       &=\sum_{i,j}\int dx_1dx_2{F'}_i(x_1){F'}_j(x_2)\hat\sigma'(x_1x_2s),
\end{split}
\label{sigscheme}
\end{equation}
where the primed quantities are associated with (\ref{dglap19}) in the
standard QCD factorization calculus. We have~\citep{qced} an attendant
shower/ME matching scheme, wherein, for example, in combining (\ref{subp15b})
with HERWIG~\citep{herwig}, PYTHIA~\citep{pythia}, MC@NLO~\citep{mcnlo}
or new shower MC's~\citep{skrzjad}, we may use either
$p_T$-matching
or shower-subtracted residuals\newline $\{\hat{\tilde{\bar\beta}}_{n,m}(k_1,\ldots,k_n;k'_1,\ldots,k'_m)\}$ to create a paradigm without double
counting that can be systematically improved order-by order in
perturbation theory -- see Refs.~\citep{qced}. \par
The stage is set for the full MC implementation of our approach. We turn next to the initial stage of this implementation -- that of the kernels in (\ref{dglap19}).\par
\section{MC Realization of IR-Improved DGLAP-CS  Theory}
In this section we describe the implementation of the 
new IR-improved kernels in the HERWIG6.5 environment, which results
in a new MC, which we denote by HERWIRI1.0, which stands for ``high energy radiation with IR improvement''\footnote{We thank M. Seymour and B. Webber for discussion on this point.}. \par
Specifically, our approach can be summarized as follows.
We modify the kernels in the HERWIG6.5 module HWBRAN and in the attendant
 related modules~\citep{bw-ms-priv} with the following substitutions:{\small
\begin{equation}\text{DGLAP-CS}\; P_{AB}  \Rightarrow \text{IR-I DGLAP-CS}\; P^{exp}_{AB}
\label{substitn}
\end{equation}}
while leaving the hard processes alone for the moment. We have in 
progress~\citep{inprog}%% (SY,BFLW,MH,SM,SJ)
the inclusion of YFS synthesized electroweak  
modules from Refs.~\citep{jad-ward}%{\Color{Magenta}Jadach et al.  MC's} 
for
HERWIG6.5, HERWIG++~\citep{herpp} hard processes, as the
CTEQ~\citep{cteq} and MRST(MSTW)~\citep{mrst} best (after 2007) parton densities
do not include the precision electroweak higher order corrections that do enter in a 1\% precison tag budget for processes such as single heavy gauge boson production in the LHC environment~\citep{radcor-ew}. 
\par
%%\end{itemize}
\def\beqa{\begin{eqnarray}}
\def\eeqa{\end{eqnarray}}
\def\beq{\begin{equation}}
\def\eeq{\end{equation}}
\def\non{\nonumber}
\def\no{\noindent }

For definiteness, let us illustrate the implementation by an example~\citep{bw-ann-rev,sjosback}, which for pedagogical reasons we will take as a simple leading
log shower component with a virtuality evolution variable, with the understanding that in HERWIG6.5 the shower development is angle ordered~\citep{bw-ann-rev} so that the evolution variable is actually $\sim E\theta$ where $\theta$ is the opening angle of the shower as defined in Ref.~\citep{bw-ann-rev} for a parton initial energy$E$. In this pedagogical example, which we take from Ref.~\citep{bw-ann-rev}, 
%\titbox{\Color{Maroon} Implementation Illustration}
the probability that no branching occurs above virtuality
cutoff $Q_0^2$ is  $\Delta_a(Q^2,Q_0^2)$ so that
%${\Color{Red}\Rightarrow}$
\beq \label{eq:splitprob}
d\Delta_a(t,Q_0^2) = \frac{-dt}{t}\Delta(t,Q_o^2)\sum_b\int dz\frac{\alpha_s}{2 \pi}P_{ba}(z),
\eeq
%\no
%${\Color{Red}\Rightarrow}$
which implies
\beq
\Delta_a(Q^2,Q_0^2)=\exp\left[ -\int_{Q_0^2}^{Q^2} \frac{dt}{t} \sum_b\int dz\frac{\alpha_s}{2 \pi}P_{ba}(z)\right].
\label{delta-a} 
\eeq
The attendant non-branching probability appearing in the evolution equation is
\flushleft{\small\beq
\Delta(Q^2,t) = \frac{\Delta_a(Q^2,Q_o^2)}{\Delta_a(t,Q_o^2)},\;t =k_a^2\equiv\;\text{virtuality of gluon $a$}.
\eeq}
The respective virtuality of parton $a$ is then generated with
\beq
\Delta_a(Q^2,t) = R,
\eeq
where $R$ is a random number uniformly distributed in $[0,1]$ .
With
\beqa
\alpha_s(Q) = \frac{2 \pi}{b_0 \log\left(\frac{Q}{\Lambda}\right)},
\eeqa
we get for example 
\beqa
\int_0^1 dz \frac{\alpha_s(Q^2)}{2 \pi} P_{qG}(z)
&=& \frac{4\pi\int_0^1 dz \frac{1}{2}\left[ z^2+(1-z)^2\right]}{2 \pi b_0\ln\left(\frac{Q^2}{\Lambda^2}\right)} \non\\ 
&=& \frac{2}{3} \frac{1}{b_0\ln\left(\frac{Q^2}{\Lambda^2}\right)}.
\eeqa
so that the subsequent integration over $dt$ yields 
%${\Color{Red}\Rightarrow}$
\beqa
&&I=\int_{Q_0^2}^{Q^2}\frac{1}{3} \frac{dt}{t}\frac{2}{ b_0 \ln\left(\frac{t}{\Lambda^2}\right)} \non \\
%,\quad t=Q^2 \non \\
&=& \frac{2}{3b_0}\ln \ln \frac{t}{\Lambda^2}|_{Q_0^2}^{Q^2} \non \\
&=& \frac{2}{3 b_0}\left[\ln \left(\frac{\ln\left(\frac{Q^2}{\Lambda^2}\right)}{\ln\left(\frac{Q_0^2}{\Lambda^2}\right)}\right)\right].
\eeqa

Finally, introducing $I$ into (\ref{delta-a}) yields 
\beqa \label{DeltaQHerwig}
\Delta_a(Q^2,Q_0^2) &=& \exp \left[-\frac{2}{3 b_0}\ln \left(\frac{\ln\left(\frac{Q^2}{\Lambda^2}\right)}{\ln\left(\frac{Q_0^2}{\Lambda^2}\right)}\right)\right]\non\\
&=& \left[\frac{\ln\left(\frac{Q^2}{\Lambda^2}\right)}{\ln\left(\frac{Q_0^2}{\Lambda^2}\right)}\right]^{-\frac{2}{3b_0}}.
\eeqa
If we now let
$\Delta_a(Q^2,t)=R$, then
\beq 
\left[\frac{\ln\left(\frac{t}{\Lambda^2}\right)}{\ln\left(\frac{Q^2}{\Lambda^2}\right)}\right]^{\frac{2}{3b_0}} = R
\eeq
which implies
\beq
t = \Lambda^2 \left(\frac{Q^2}{\Lambda^2}\right)^{R^{\frac{3 b_0}{2}}}.
\label{t-herwig}
\eeq
Recall in HERWIG6.5~\citep{herwig} we have 
\beqa
b_0 &=& \left(\frac{11}{3}n_c - \frac{2}{3}n_f\right) \non \\ 
&=& \frac{1}{3}\left(11n_c - 10 \right), \quad n_f =5 \non \\
&\equiv& \frac{2}{3} \bf{BETAF}.
\eeqa
where in the last line we used the notation in HERWIG6.5.
The momentum available after a  $q\bar{q}$ split in HERWIG6.5~\citep{herwig} 
is given by
\beq
QQBAR = QCDL3\left(\frac{QLST}{QCDL3}\right)^{R^{BETAF}},
\eeq
in complete agreement with (\ref{t-herwig}) when we note the
identifications $t=QQBAR^2,\;\Lambda\equiv QCDL3,\; Q\equiv QLST$.\par
The leading log exercise leads to the same algebraic relationship that
HERWIG6.5 has between $QQBAR$ and $QLST$ but 
we stress that in HERWIG6.5
these quantities are the angle-ordered counterparts of the virtualities 
we used in our example, so that the shower is angle-ordered.\par
\begin{figure}[h]
			\begin{center}
				\scalebox{0.45}{\includegraphics{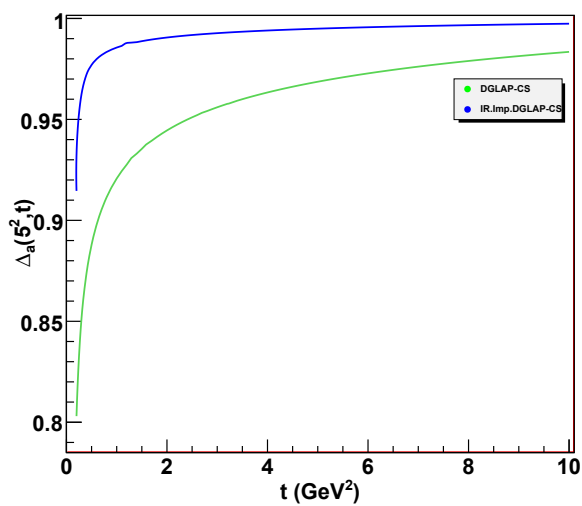}}
			\end{center}
\caption{Graph of $\Delta_a(Q^2,t)$ for the DGLAP-CS and IR.Imp.DGLAP-CS kernels (\ref{DeltaQHerwig}, \ref{DeltaQWard}). Q$^2$ is a typical virtuality closer to the squared scale of the hard sub-process -- here we use $Q^2=25$GeV$^2$ for illustration.} 
\label{iri-vs-cs}
\end{figure}

  
When we repeat the above calculation for the IR-Improved kernels in (\ref{dglap19}), we have
\beq
P_{qG}^{exp}(z) = F_{YFS}(\gamma_G)e^{\frac{1}{2}\delta_G}\frac{1}{2} 
                    \biggl[ z^2(1-z)^{\gamma_G} + (1-z)^2z^{\gamma_G} \biggr]
\eeq
so that
\beqa
\int_0^1 dz \frac{\alpha_s\left(Q^2\right)}{2 \pi} P_{qG}(z)^{exp}
 &= \frac{4F_{YFS}(\gamma_G)e^{\frac{1}{2}\delta_G} }{b_0 \ln\left(\frac{Q^2}{\Lambda^2}\right)\left(\gamma_G+1\right)\left(\gamma_G+2\right)}\non\\
&\qquad\qquad \frac{1}{\left(\gamma_G+3\right)}.
\eeqa
This leads to the following integral over $dt$
\beqa
I&=\int_{Q_0^2}^{Q^2}\frac{dt}{t} \frac{4F_{YFS}(\gamma_G)e^{\frac{1}{2}\delta_G} }{b_0 \ln\left(\frac{t}{\Lambda^2}\right)\left(\gamma_G+1\right)\left(\gamma_G+2\right)\left(\gamma_G+3\right)} \non\\
%,\quad t=Q^2 \non \\ 
&=\frac{4 F_{YFS}(\gamma_G)e^{0.25 \gamma_G}}{b_0\left(\gamma_G+1\right)\left(\gamma_G+2\right)\left(\gamma_G+3\right)}\non\\
&\qquad Ei\left(1,\frac{8.369604402}{b_0\ln\left(\frac{t}{\Lambda^2}\right)}\right) \Bigg\vert_{Q_0^2}^{Q^2}.
\eeqa
Here we have used
\beq
\delta_G=\frac{\gamma_G}{2} + \frac{\alpha_s C_G}{\pi}\left(\frac{\pi^2}{3} - \frac{1}{2}\right).
\eeq
%with $C_G=3$ the gluon quadratic Casimir invariant.
We finally get the IR-improved formula
\beq \label{DeltaQWard}
\Delta_a(Q^2,t) = \exp\left[-\left(F\left(Q^2\right)-F\left(t\right)\right)\right],
\eeq
where{\small
\beq
\begin{split}
F(Q^2) &= \frac{4 F_{YFS}(\gamma_G)e^{0.25 \gamma_G}}{b_0\left(\gamma_G+1\right)\left(\gamma_G+2\right)\left(\gamma_G+3\right)} \\
&\qquad Ei\left(1,\frac{8.369604402}{b_0\ln\left(\frac{Q^2}{\Lambda^2}\right)}\right),
\end{split}
\eeq}
and $Ei$  is the exponential integral function.
In Fig.~\ref{iri-vs-cs} we show the difference between the two results for 
$\Delta_a(Q^2,t)$. We see that they agree within a few \% except for the softer values of $t$, as expected. We look forward to determining definitively
whether the experimental data prefer one over the other. This detailed study will appear elsewhere~\citep{elswh} but we begin the discussion below with a view on recent FNAL data. 
Again, we note that the comparison in Fig.~\ref{iri-vs-cs} is done here at the leading log virtuality level, but the sub-leading effects we have suppressed in discussing it will not change our general conclusions drawn therefrom.
\par

We have carried out the corresponding changes for all of the kernels in (\ref{dglap19}) in the HERWIG6.5 environment, with its angle-ordered showers, 
resulting in the new MC, HERWIRI1.0,
in which the ISR parton showers have IR-improvement as given by
the kernels in (\ref{substitn}). In the original 
release of the program, v. 1.0, 
we stated that the time-like parton showers had 
been completely IR-improved in a way that suggested the space-like parton 
showers had not yet been IR-improved at all. Then, in release 1.02,
we stated that 
the only part of the 
space-like parton showers without IR-improvement in v. 1.0
is that associated with 
HERWIG6.5's space-like module HWSGQQ for the space-like branching process
$G\rightarrow q\bar{q}$, a process which is not IR divergent and which is, 
in any case, a sub-dominant part of the shower. 
The module HWSQGG was IR-improved as well in the release HERWIRI1.02.
This was in fact an oversight, as the module HWSFBR which controls
the remainder of the space-like branching processes was also still in need
of IR-improvement in versions 1.0 and 1.02\footnote{We thank Profs. B. Webber and M. Seymour for discussion here.}. We have done the required IR-improvement
of the latter module as well in version 1.031. While the effect
in going from version 1.0 to version 1.02 is small, that in going from
version 1.02 to 1.031 is not in general and we recommend
version 1.031 for the best precision.
%The effect is small, as these considerations suggest: we see effects at a 
%level comparable to the errors on the MC data 
%in our plots when going from version 1.0 to 1.02.
We now illustrate some of the results we have obtained in comparing 
ISR showers in HERWIG6.5 and with those in HERWIRI1.0(31) at LHC
%%to be added when FNAL figs ready: 
and at FNAL
energies, where some comparison with real data is also featured at the FNAL
energy.
%\titbox{\Color{Maroon}RESULTS}
Specifically, we compare the z-distributions, $p_T$-distributions, etc., that result from the IR-improved and usual
DGLAP-CS showers in what follows\footnote{Note that similar results for PYTHIA and MC@NLO are in progress in general; for MC@NLO we have some
initial results already in particular cases -- see the discussion below.}.\par
\begin{figure}[h]
\begin{center}
\includegraphics[height=80mm,angle=90]{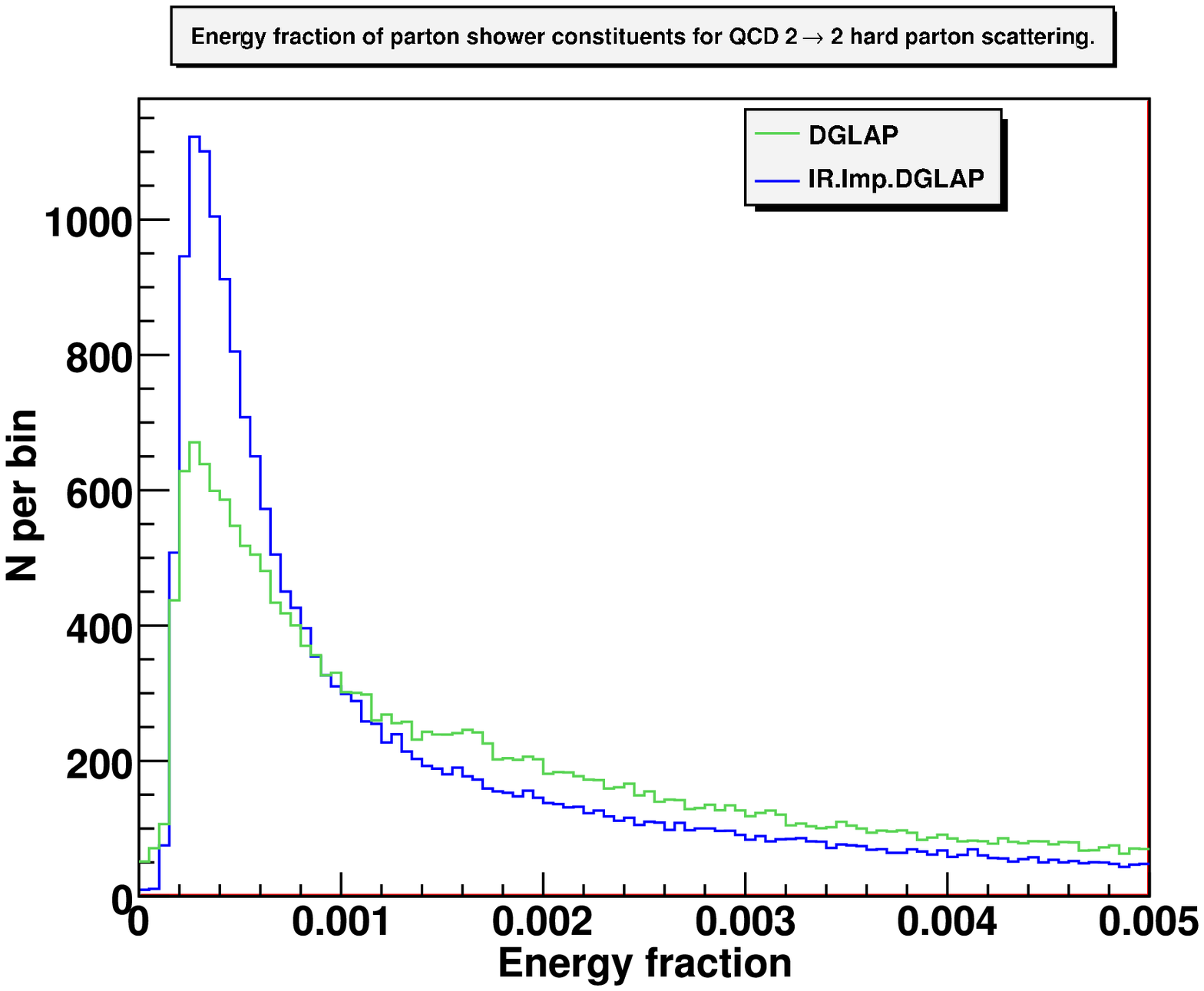}
\end{center}
\caption{ The z-distribution(ISR parton energy fraction) shower comparison in HERWIG6.5.}
\label{fighw1}
\end{figure}
\begin{figure}[h]
\begin{center}
\includegraphics[height=75mm]{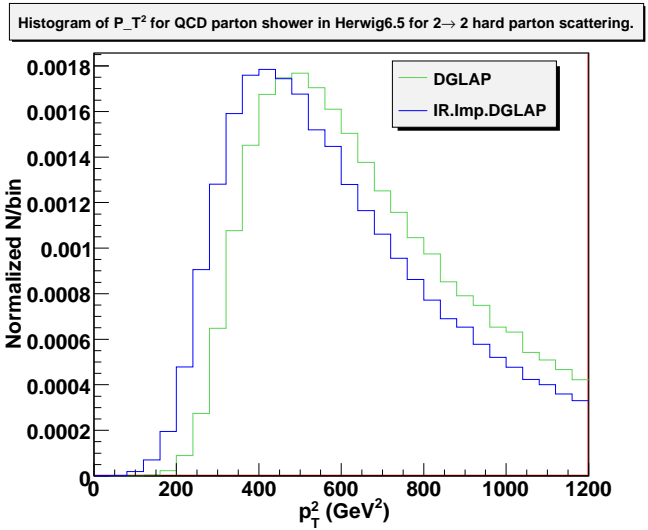}
\end{center}
\caption{The  $P_T^2$-distribution (ISR parton) shower comparison in HERWIG6.5.}
\label{fighw2}
\end{figure}
\begin{figure}[h]
\begin{center}
\includegraphics[height=70mm]{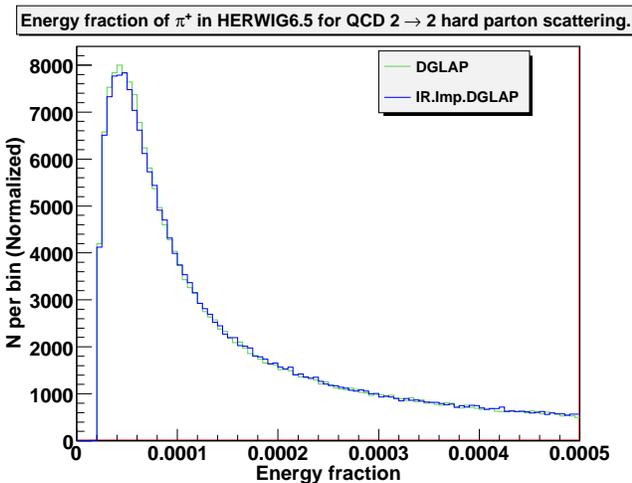}
\end{center}
\caption{The $\pi^+$ energy fraction distribution shower comparison in HERWIG6.5.}
\label{fighw3}
\end{figure}
\begin{figure}[h]
\begin{center}
\includegraphics[height=70mm]{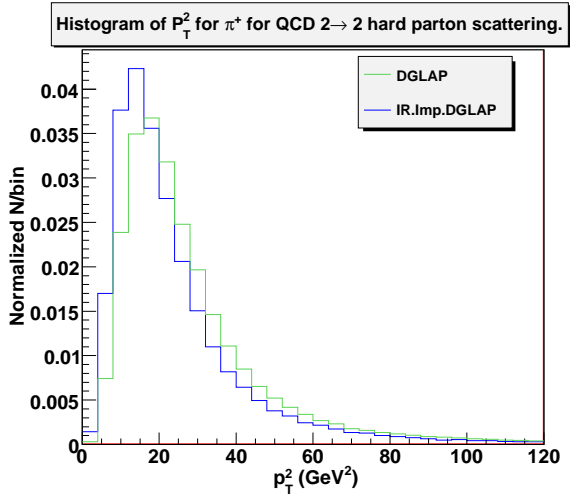}
\end{center}
\caption{The $\pi^+$ $P_T^2$-distribution shower comparison in HERWIG6.5.}
\label{fighw4}
\end{figure}

First, for the generic 2$\rightarrow$2 hard processes at LHC energies (14 TeV) we get the comparison shown Figs.~\ref{fighw1}, \ref{fighw2} for the respective ISR $z$-distribution and $p_T^2$ distribution at the parton level. 
Here, there are no cuts placed on the MC data and we define $z$ as
$z=E_{\text{parton}}/E_{\text{beam}}$ where $E_{\text{beam}}$ is the cms beam energy and $E_{\text{parton}}$ is the respective parton energy in the cms system. The two quantities $z$ and  $p_T^2$ for partons are of course not directly observable but their distributions show the softening of the IR divergence as we expect.
%%%START-HERE
%%\ref{fighw1},\ref{fighw2},\ref{fighw3}:
%\ref{fighw1},\ref{fighw2},\ref{fighw3}:
Turning next to the similar quantities for the $\pi^+$ production in the 
generic $2 \rightarrow 2$ hard processes at LHC, we see in Figs.~\ref{fighw3}, \ref{fighw4} that spectra in the former are similar and spectra in the latter 
are again softer in the IR-improved case. These spectra of course would be 
subject to some ``tuning'' in a real experiment and we await with anticipation the outcome of such 
an effort in comparison to LHC data.\par
\begin{figure}[h]
\begin{center}
\includegraphics[height=70mm]{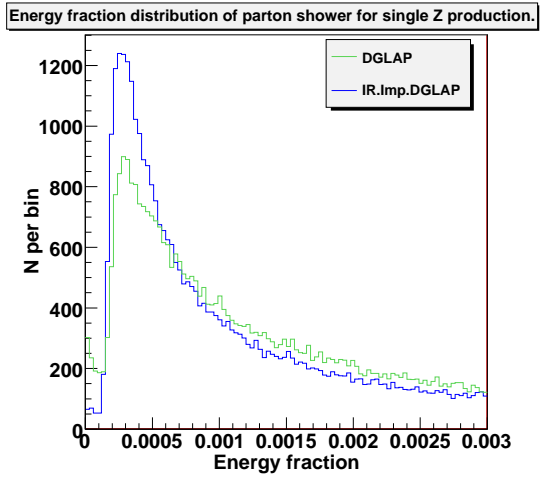}
\end{center}
\caption{The $z$-distribution(ISR parton energy fraction) shower comparison in HERWIG6.5.}
\label{fighw5}
\end{figure}

\begin{figure}[h]
\begin{center}
\includegraphics[height=70mm]{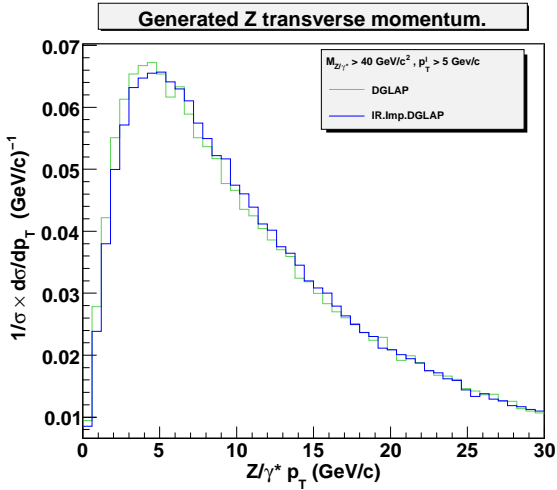}
\end{center}
\caption{The $Z$ p$_T$-distribution(ISR parton shower effect) comparison in HERWIG6.5.}
\label{fighw6}
\end{figure}

\begin{figure}[h]
\begin{center}
\includegraphics[height=70mm]{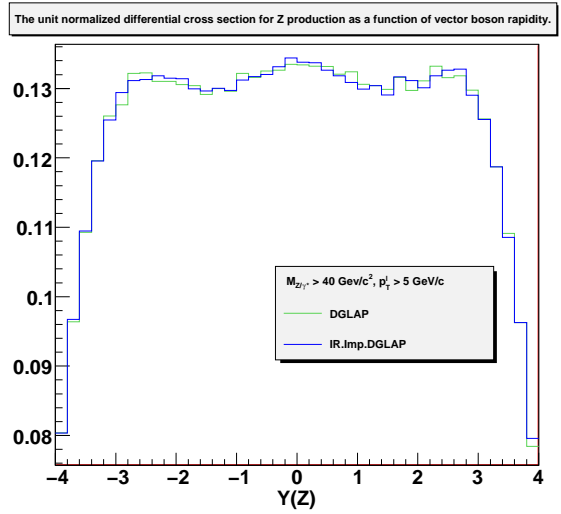}
\end{center}
\caption{The $Z$ rapidity-distribution(ISR parton shower) comparison in HERWIG6.5.}
\label{fighw7}
\end{figure}

We turn next to the luminosity process of single $Z$ production at the LHC, where in Figs.~\ref{fighw5},\ref{fighw6},\ref{fighw7} we show respectively the ISR parton energy fraction distribution, the Z-p$_T$ distribution, and the Z-rapidity distribution
with cuts on the acceptance as $40\text{GeV}<M_Z,\; p^\ell_T>5\text{GeV}$
%,\; |\eta_\ell|<50$ 
for $Z\rightarrow \mu\bar{\mu}$ -- all lepton rapidities are
included. For the energy fraction distribution and the p$_T$ distributions we again see softer spectra in the former and we see similar spectra in the latter in the IR-improved case. For the rapidity plot, we see the migration of some events to the higher values of $|\eta|$, which is not 
inconsistent with a softer spectrum for the IR-improved case
\footnote{One might wonder why we show the $Z$ rapidity here as the soft gluons
which we study only have an indirect affect on it via momentum conservation?
But, this means that the rapidity predicted by the IR-improved showers
should be close to that predicted by the un-improved showers 
and we show this cross-check is
indeed fulfilled in our plots.}. 
We look forward to the confrontation with experiment,
where again we stress that in a real experiment, a certain amount of ``tuning'' with affect these results. The question will always be which set of distributions gives a better $\chi^2$ per degree of freedom.\par
%\item Single Z-production at LHC
\begin{figure*}[t]
%\begin{center}
%%%%\epsfig{file=fig01.ps}
%\epsfig{file=en_sgam-200.eps,width=140mm,height=130mm}
%\end{center}
%\vspace{ -8mm}
%\baselineskip=7mm
\centering
\setlength{\unitlength}{0.1mm}
%%%%%%%%%%%%%%%%%%%%%%%%%%%%%%%%%%
%%%\begin{picture}(1600, 1540)
\begin{picture}(1600, 720)
%%\put( 450, 1530){\makebox(0,0)[cb]{\bf (a)} }
%%\put(1230, 1530){\makebox(0,0)[cb]{\bf (b)} }
%%\put(   0, 870){\makebox(0,0)[lb]{\epsfig{file=ef.ps,angle=270,
%%                                        width=80mm}}}
%%\put( 800, 870){\makebox(0,0)[lb]{\epsfig{file=hwptc.ps,angle=270,
%%                                        width=80mm}}}
%%%\put( 250, 660){\makebox(0,0)[cb]{\bf (a)} }
%%%\put(800, 660){\makebox(0,0)[cb]{\bf (b)} }
%%%\put(   0, 100){\makebox(0,0)[lb]{\epsfig{file=vgcut.eps,angle=90,
%%%                                        width=55mm}}}
%%%\put( 550, 100){\makebox(0,0)[lb]{\epsfig{file=vgcutir.eps,angle=90,
%%%                                        width=55mm}}}
\put( 350, 690){\makebox(0,0)[cb]{\bf (a)} }
\put(1130, 690){\makebox(0,0)[cb]{\bf (b)} }
\put(   0, 50){\makebox(0,0)[lb]{\includegraphics[width=65mm]{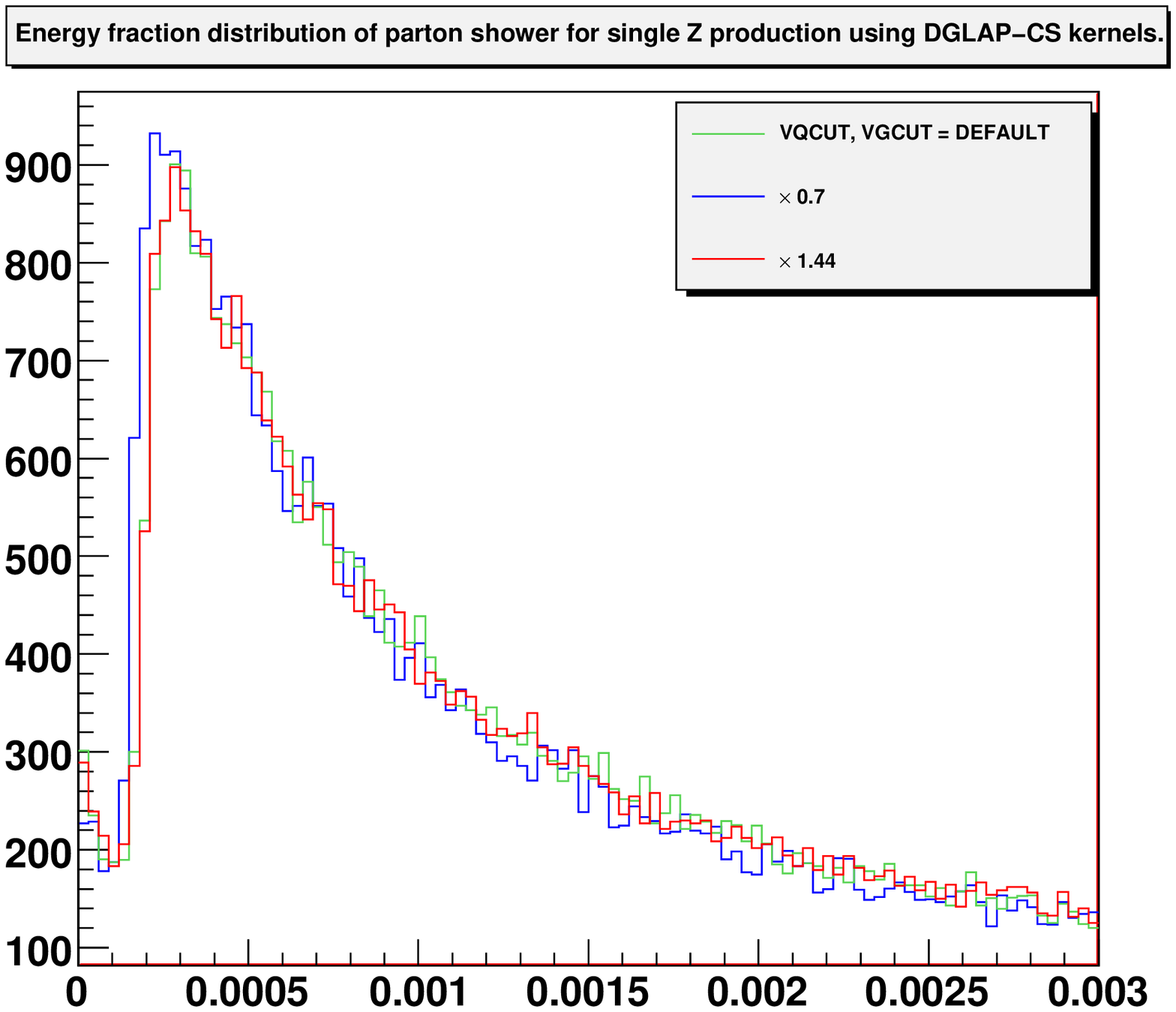}}}
%{\epsfig{file=endglap.ps,angle=90,
%                                        width=80mm}}}
%\put( 800, 100){\makebox(0,0)[lb]{\includegraphics[width=80mm,angle=90]{endglapir.ps}}}
\put( 800, 50){\makebox(0,0)[lb]{\includegraphics[width=65mm]{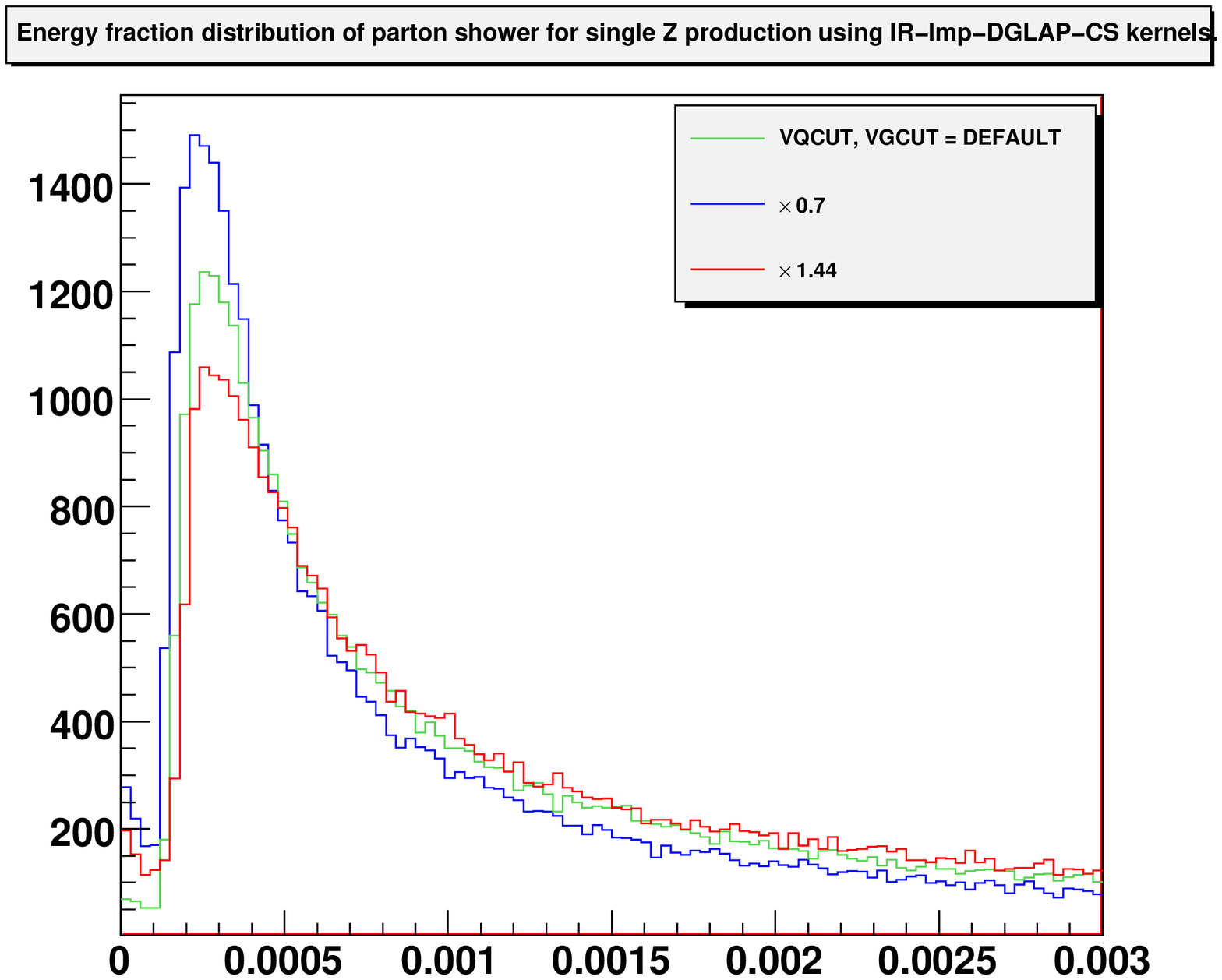}}}
%{\epsfig{file=endglapir.ps,angle=90,
%                                        width=80mm}}}
\end{picture}
%\vspace{ -1.5mm}
\caption{IR-cut-off sensitivity in z-distributions of the ISR parton energy fraction: (a), DGLAP-CS 
(b), IR-I DGLAP-CS -- for the single Z hard
sub-process in HERWIG-6.5 environment.
%\centerline{\Color{PineGreen}COMPARISON WITH DATA IMMINENT.} 
}
\label{fighw8}
\end{figure*}

Finally, we turn the issue of the IR-cut-off in HERWIG6.5. In HERWIG6.5,
there is are IR-cut-off parameters used to separate real and virtual effects
and necessitated by the +-function representation of the usual 
DGLAP-CS kernels. In HERWIRI, these parameters can be taken arbitrarily close to zero, as the IR-improved kernels are integrable~\citep{irdglap1,irdglap2}. We now illustrate the difference in IR-cut-off response by comparing it for HERWIG6.5 and HERWIRI:
we change the default values of the parameters in HERWIG6.5 by factors of .7 and 1.44 as shown in the Fig.~\ref{fighw8}. We see that the harder cut-off reduces the phase space only significantly for the IR-improved kernels and that the softer cut-off has also a small effect on the usual kernels spectra whereas as expected
the IR-improved kernels spectra move significantly toward softer values as a convergent integral
would lead one to expect. This should lead to a better description of the soft
radiation data at LHC. We await confrontation with experiment accordingly.
\par

\begin{figure*}[t]
%\begin{center}
%%%%\epsfig{file=fig01.ps}
%\epsfig{file=en_sgam-200.eps,width=140mm,height=130mm}
%\end{center}
%\vspace{ -8mm}
%\baselineskip=7mm
\centering
\setlength{\unitlength}{0.1mm}
%%%%%%%%%%%%%%%%%%%%%%%%%%%%%%%%%%
%%%\begin{picture}(1600, 1540)
\begin{picture}(1600, 720)
%%\put( 450, 1530){\makebox(0,0)[cb]{\bf (a)} }
%%\put(1230, 1530){\makebox(0,0)[cb]{\bf (b)} }
%%\put(   0, 870){\makebox(0,0)[lb]{\epsfig{file=ef.ps,angle=270,
%%                                        width=80mm}}}
%%\put( 800, 870){\makebox(0,0)[lb]{\epsfig{file=hwptc.ps,angle=270,
%%                                        width=80mm}}}
%%%\put( 250, 660){\makebox(0,0)[cb]{\bf (a)} }
%%%\put(800, 660){\makebox(0,0)[cb]{\bf (b)} }
%%%\put(   0, 100){\makebox(0,0)[lb]{\epsfig{file=vgcut.eps,angle=90,
%%%                                        width=55mm}}}
%%%\put( 550, 100){\makebox(0,0)[lb]{\epsfig{file=vgcutir.eps,angle=90,
%%%                                        width=55mm}}}
\put( 350, 690){\makebox(0,0)[cb]{\bf (a)} }
\put(1130, 690){\makebox(0,0)[cb]{\bf (b)} }
%\put(   0, 100){\makebox(0,0)[lb]{\includegraphics[width=80mm,angle=90]{rapcdf1.ps}}}
\put(   0, 50){\makebox(0,0)[lb]{\includegraphics[width=70mm]{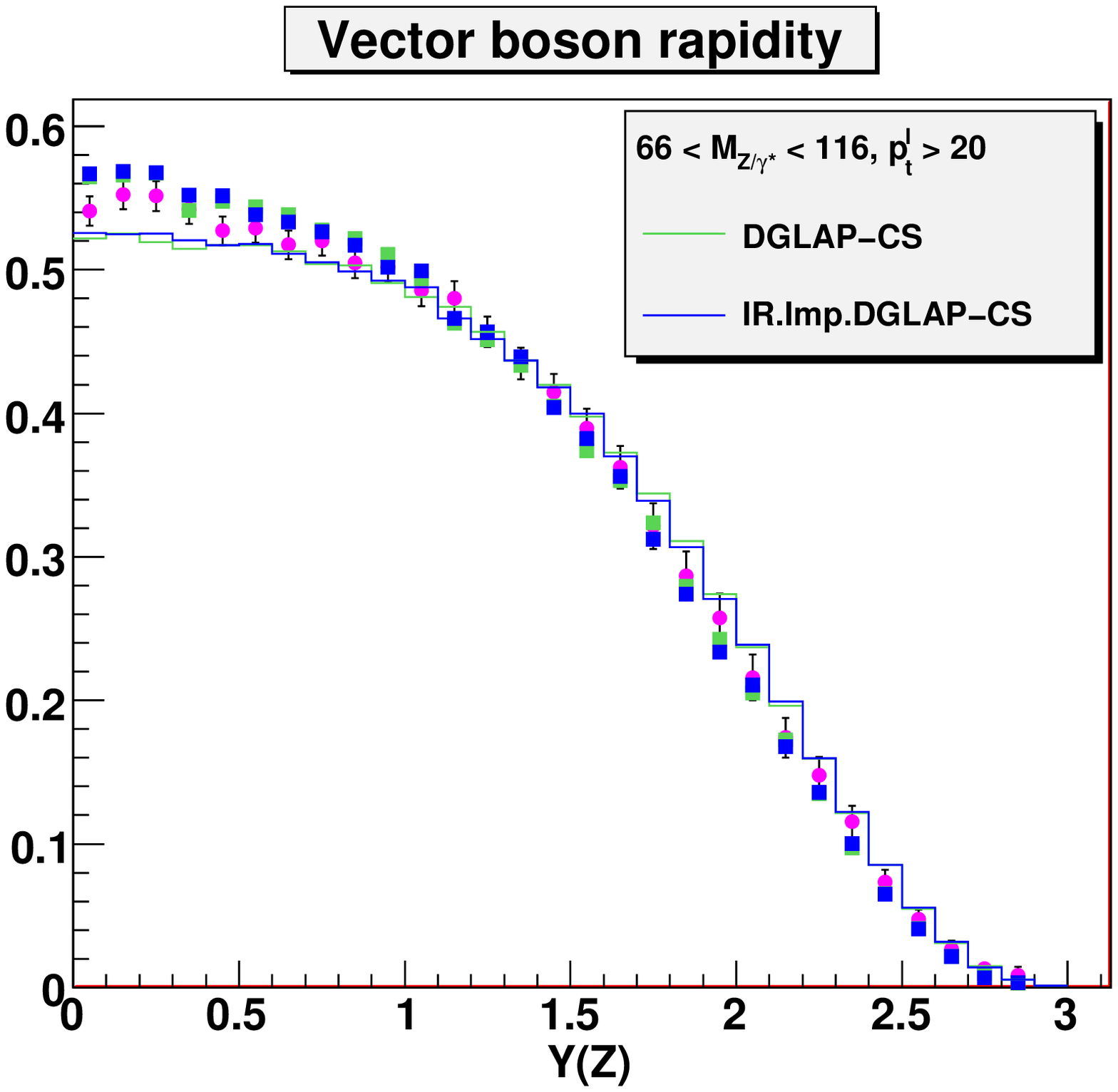}}}
%{\epsfig{file=rapcdf1.ps,angle=90,
%                                        width=80mm}}}
%\put( 800, 100){\makebox(0,0)[lb]{\includegraphics[width=80mm,angle=90]{ptplot090609.ps}}}
\put( 800, 50){\makebox(0,0)[lb]{\includegraphics[width=70mm]{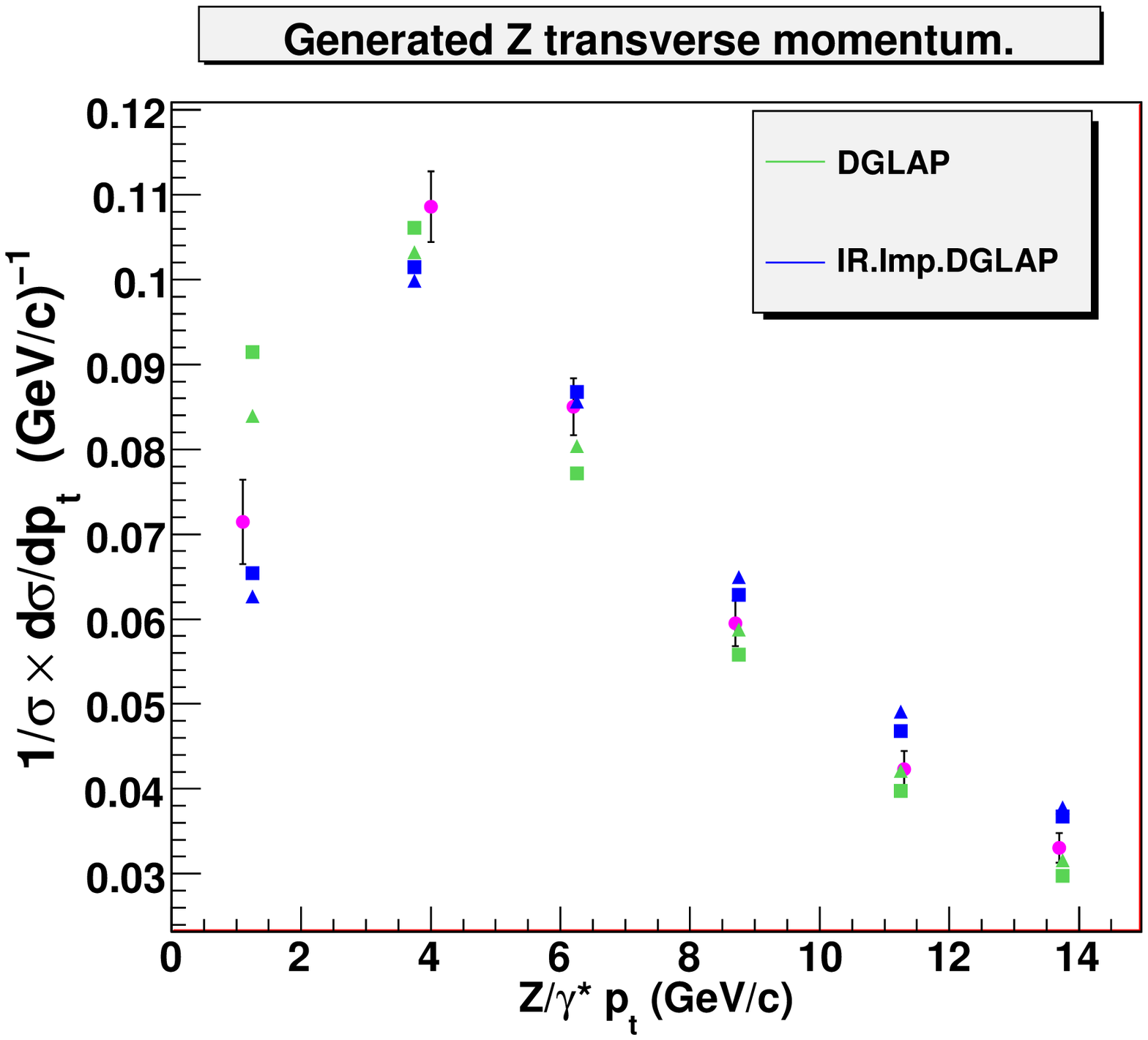}}}
%{\epsfig{file=ptplot090609.ps,angle=90,
%                                        width=80mm}}}
\end{picture}
%\vspace{ -1.5mm}
\caption{Comparison with FNAL data: (a), CDF rapidity data on
($Z/\gamma^*$) production to $e^+e^-$ pairs, the circular dots are the data; 
(b), D0 $p_T$ spectrum data on ($Z/\gamma^*$) production to $e^+e^-$ pairs,
the circular dots are the data, the blue triangles are HERWIRI1.031, the green triangles are HERWIG6.510 -- in both (a) and (b) the blue squares are MC@NLO/HERWIRI1.031, and the green squares are MC@NLO/HERWIG6.510. These are untuned theoretical results..
%\centerline{\Color{PineGreen}COMPARISON WITH DATA IMMINENT.} 
}
\label{fighw9}
\end{figure*}

We finish this initial comparison discussion by turning to the data from 
FNAL on the Z rapidity and $p_T$ spectra as reported in 
Refs.~\citep{galea,d0pt}. We show these results, for 1.96TeV cms 
energy, in Fig.~\ref{fighw9}. We see that, in the case of the 
CDF rapidity data, HERWIRI1.0(31)
and HERWIG6.5 both give a reasonable 
overall representation of the data but that
HERWIRI1.0(31) is somewhat closer to the data for small values of $Y$. The two $\chi^2$/d.o.f are 1.77 and 1.54
for HERWIG6.5 and HERWIRI1.031 respectively. The data 
errors in Fig.~\ref{fighw9}(a)
do not include luminosity and PDF errors~\citep{galea}, so that
they can only be used conditionally at this point.
Including the NLO
contributions to the hard process via MC@NLO/HERWIG6.510
and MC@NLO/HERWIRI1.031\citep{mcnlo}\footnote{We thank S. Frixione for
helpful discussions with this implementation.} improves the agreement for both
HERWIG6.5 and for HERWIRI1.031, where the $\chi^2$/d.o.f are changed
to 1.40 and 1.42 respectively.
%The two $\chi^2$/p.d.f. are xxx and xxx for HERWIRI1.0 and HERWIG6.5 respectively. 
%The two
%respective $\chi^2$/d.o.f. are xxx and yyy for  HERWIG6.5 and HERWIRI1.0
%for all the data in Fig.~\ref{fighw9}(b).
%The two attendant
%$\chi^2$/p.d.f. are xxx and xxx for HERWIRI1.0 and HERWIG6.5 respectively.
\par
%\begin{figure}
%\begin{center}
%\epsfig{file=vguct.ps,height=60mm}
%\end{center}
%\caption{\baselineskip=7mm     The z-distribution(ISR parton energy fraction) shower comparison in HERWIG6.5 -- preliminary results.}
%\label{hw1}
%\end{figure}
%{\Color{Brown} SIMILAR RESULTS FOR PYTHIA and MC@NLO IN PROGRESS.}
%\item {\Color{PineGreen}COMPARISON WITH DATA IMMINENT.}
%}
%\end{itemize}
\noindent
For the D0 $p_T$ data, we see that HERWIRI1.0(31)
 gives a better fit to the data
compared to HERWIG6.5 for low $p_T$, 
(for $p_T<12.5$GeV, the $\chi^2$/d.o.f. are
$\sim$ 2.5 and 3.3 respectively if we add the statistical and systematic
errors), showing that the IR-improvement makes a better representation
of QCD in the soft 
regime for a given fixed order in perturbation theory.
Including the results of MC@NLO~\citep{mcnlo}
%\footnote{We thank S. Frixione for helpful discussion on this implementation.} 
improves the $\chi^2$/d.o.f for
the HERWIRI1.031 in both the soft and hard regimes and it improves
the HERWIG6.510 $\chi^2$/d.o.f for $p_T$ near $3.75$ GeV
where the distribution peaks. For $p_T<7.5$GeV the $\chi^2$/d.o.f for
the MC@NLO/HERWIRI1.031 is 1.5 whereas that for MC@NLO/HERWIG6.510 is 
worse. 
%For the higher values of $p_T$ both results show the need for the
%exact higher order corrections as expected, but HERWIG6.5 is
%closer to the data than is HERWIRI1.0 as we see already 
\par
\section{Conclusions}\label{concl}
Our new MC HERWIRI1.0(31)
sets the stage for
the further implementation
of the attendant~\citep{qced} 
new approach to precision QED$\times$QCD predictions for LHC physics
by the introduction of the
respective resummed residuals needed to systematically improve
the precision tag to the 1\% regime for such processes as single heavy gauge boson production, for example. Here, we already note that this new
IR-improved MC, HERWIRI1.0(31), available at http://thep03.baylor.edu, is expected to allow for a better $\chi^2$ per degree of freedom in data analysis of high energy
hadron-hadron scattering for soft radiative effects and we have given 
evidence that this is indeed the case.
\begin{acknowledgments}
%\end{acknowledgments}
%\section*{Acknowledgments}
%The authors wish to thank JACoW for their guidance in preparing
%this template.
%
%Work supported by Department of Energy contract DE-AC02-76SF00515.
%\end{acknowledgments}
One of us (B.F.L.W) acknowledges helpful discussions with Prof. Bryan Webber
and Prof. M. Seymour. B.F.L. Ward also thanks Prof. L. Alvarez-Gaume and Prof. W. Hollik for the support and kind hospitality of the CERN TH Division and of the Werner-Heisenberg Institut, MPI, Munich, respectively, while this work was in progress.

Work partly supported by US DOE grant DE-FG02-05ER41399 and 
% the Polish Government
%grants KBN 2P30225206 and 2P03B17210, the Maria Sk\l{}odowska-Curie
%Joint Fund II PAA/DOE-97-316, and
by NATO Grant PST.CLG.980342.
\end{acknowledgments}

%%%STARTHERE
\end{document}